\documentclass[a4paper]{article}

\usepackage{icrc2013}

\title{Detection prospects for short time-scale transient events at VHE with current and next generation Cherenkov observatories}

\shorttitle{Detection prospects for transient events at VHE}

\authors{
S.~Lombardi$^{1,2}$,
A.~Carosi$^{1,2}$,
L.A.~Antonelli$^{1,2}$
}

\afiliations{
$^1$ INAF--Osservatorio Astronomico di Roma, Via Frascati 33, 00040 Monte Porzio Catone (RM), Italy\\
$^2$ ASI Science Data Center, Via del Politecnico, 00133 Roma, Italy \\
}

\email{saverio.lombardi@oa-roma.inaf.it}

\abstract{
In the current view of Gamma-Ray Burst (GRB) phenomena, an emission component extending up 
to the very-high energy (VHE, $E>30$~GeV) domain is though to be a relatively common feature 
at least in the brightest events. This leads to an unexpected richness of possible theoretical 
models able to describe such phenomenology. Hints of emission at tens of GeV are indeed known 
since the EGRET observations during the '90s and confirmed in the \emph{Fermi}--LAT data. 
However, our comprehension of these phenomena is still far to be satisfactory. In this respect, the VHE 
characterization of GRBs may constitute a breakthrough for understanding their physics and, 
possibly, for providing decisive clues for the discrimination among different proposed emission 
mechanisms, which are barely distinguishable at lower energies. The current generation of 
Cherenkov observatories, such as the MAGIC telescopes, have opened the possibility to extend 
the measurement of GRB emission, and in general to any short time-scale transient phenomena, 
from few tens of GeV up to the TeV energy range, with a higher sensitivity with respect to 
$\gamma$-ray space-based instruments. In the near future, a crucial role for the VHE observations of GRBs 
will be played by the Cherenkov Telescope Array (CTA), thanks to its about 
one order of magnitude better sensitivity and lower energy threshold with respect to current 
instruments. In this contribution, we present a method aimed at providing VHE detection 
prospects for observations of GRB-like transient events with Cherenkov telescopes. In particular, 
we consider the observation of the transient event GRB~090102 as a test case for the method and 
show the achieved detection prospects under different observational conditions for the MAGIC 
telescopes and CTA.
}


\begin{document}
\maketitle

\section{Introduction}
\label{sec1}
Time domain astrophysics is going to play a key role in our understanding of different kind
of cosmic sources. In particular, the discovery of high-energy $\gamma$-rays from an 
unexpected large variety of transient events with time-scale ranging from millisecond up to 
days poses a new series of theoretical problems~\cite{ge12a}. The list of $\gamma$-ray band 
transient sources comprises both local phenomena, as terrestrial and solar $\gamma$-ray flares, 
as well as galactic and extra-galactic transient events. Furthermore, short time-scale variability
has long been observed in active galaxies, especially for blazars-class objects \cite{ne13}. 
The extension, when possible, of the multi-wavelength coverage up to very high energy (VHE, $E>30$~GeV)
can provide powerful diagnostic tools to understand the nature of these objects and discriminate among the 
different proposed interpretative scenarios. In particular, Gamma-Ray Bursts (GRBs) have long 
been seen as the transient events per excellence. At their peak activity, GRBs become the most 
luminous objects of the Universe releasing enormous amounts of energy from $10^{52}$\,erg to 
$10^{54}$\,erg of isotropic-equivalent energy over brief periods of $0.01$~--~$1000$~s. 
They usually show their phenomenology mainly in the $10$\,keV~--~$1$\,MeV energy band with 
extremely rapid and irregular variability (see e.g.~\cite{ge12b} for a review). However, recent 
results from the \emph{Fermi}--LAT (Large Area Telescope) have showed that, at least for the brightest events, a GeV emission 
from GRBs is a relatively common phenomenon~\cite{gr10}. Interestingly, in the majority of the 
LAT GRBs, GeV emission occurs with a significant delay with respect to the MeV and sub-MeV 
emission and it lasts longer than the emission detected by the \emph{Fermi}--GBM (Gamma-ray Burst 
Monitor). 

While sub-MeV GRB emission can often be explained by electron synchrotron processes, the 
theoretical framework of a possible second emission component in the high-energy regime is 
still less clear. Furthermore, the recent detection of GRB~130427A with photons up to $94$\,GeV~\cite{zh13}
also indicates the possibility of an observable VHE counterpart. Several authors have derived
predictions for the VHE emission from the GRB prompt and afterglow phase taking into account 
non-thermal leptonic and hadronic processes~\cite{zh01} as well as photospheric up-scattered 
emission~\cite{to11}. 
VHE observations with sufficiently high sensitivity may definitively 
solve, or at least strongly constrain, the mechanisms for prompt and early afterglow emission 
through broader energy band coverage. At the same time, and also for other GRB-like transient 
phenomena, such as Tidal Disruption Events~\cite{al13b}, VHE data may throw light on some physic
aspects which are still poorly understood, including the determination of the bulk Lorentz 
factor of the outflow, the dynamics of particle acceleration, and the jet formation. 

The capabilities of a significant detection at VHE from such a kind of events are strongly 
related, on the one hand, to the scientific performance of the instrument, and, on the other hand, 
to the short-time scale features of the $\gamma$-ray signal. The strong time dependence 
of the emission hence makes the usual considerations based on the sensitivity of the instrument
not suitable for providing detection prospects for transient events at VHE.

In this work, we illustrate a method for evaluating the detectability of GRB-like events with
the MAGIC stereoscopic system and the next generation Cherenkov Telescope Array (CTA) based on the 
time evolution of the significance of the observation.
As a test case, we consider the particular event GRB~090102~\cite{ma09} (which was observed
by the MAGIC-I telescope~\cite{al13a}) and show its detection prospect results.

\section{Observations of GRB-like events with current and next generation Cherenkov telescopes}
\label{sec2}
Despite the remarkable results of the \emph{Fermi}--LAT, the number of detected photons above few 
tens of GeV remains rather limited, motivating follow-up observations with much better 
sensitivity in the VHE band with the use of Imaging Atmospheric Cherenkov Telescopes (IACTs)~\cite{vo09}. 
Thanks to the technical evolution of such kind of instruments, in the last decade, intense 
studies have been performed on GRB science with IACTs to explore possible VHE band emission 
for these enigmatic events. 

Since already several years, current IACTs, such as MAGIC\cite{magic}, H.E.S.S.\cite{hess}, 
and VERITAS\cite{veritas}, despite their reduced duty cycle, started observational programs on 
GRB follow-up, making the $\sim 100$\,GeV~--~TeV energy range accessible to GRB observations.
As a matter of fact, several attempts to observe GRB emission have been reported by current IACT 
collaborations (e.g.~\cite{al10,ah09,ac11}). In all cases only upper limits have been derived. 
However, it is well known that the flux above $\sim 100$\,GeV is affected by the attenuation by 
pair production with the lower energetic (optical/IR) photons of the diffuse Extragalactic 
Background Light (EBL)~\cite{do11}. The consequent Universe opacity heavily affect Cherenkov observations, 
almost hindering the detection for relatively high redshift ($z>0.5$) sources. This is the case 
for GRBs which have long been known to have redshift slightly larger than 2. This basically
implies that the expected detection rate for current Cherenkov telescopes is estimated to be around 
$0.1$~--~$0.2$~GRBs/year and should significantly improve only with the coming CTA, for which 
GRBs will be among primary targets~\cite{in13}. 

The CTA project~\cite{act11} aims at developing the next generation ground-based instrument 
dedicated to the observations in the VHE $\gamma$-ray band. In the current layout of CTA, 
the arrays will consist of three types of telescopes with different main mirror sizes in order to cover 
the full energy range from few tens of GeV up to a hundred of TeV. The lowest energy band (i.e., where 
GRBs are mainly foreseen to show their activity) will be covered by few 24-m Large Size Telescopes (LSTs). 
With respect to current IACT facilities, CTA will mainly benefit from a lower energy threshold 
(down to $\sim 20$ GeV), a much larger effective collection area, particularly in the few tens of GeV energy range,  
and a sensitivity about one order of magnitude better in the whole energy range~\cite{al12,be13}.
Furthermore, LSTs are conceived to have rapid slewing capability with a repositioning 
time of around $180^{\circ}$ azimuthal rotation in $20$\,s (i.e. comparable to the 
performance achieved by the MAGIC telescopes~\cite{ga10}). In some cases, this will permit GRB 
observations during prompt emission phase while the majority of the events can be observed at 
early afterglow stage. Estimate based on different LST performance and GRB statistics currently 
foresees a still limited detection rate of few bursts per year~\cite{in13}. However, CTA high
sensitivity will permit the collection of enough VHE photons to perform time-resolved studies 
of the observed events.

\section{Detection prospects for transient events at VHE}
\label{sec3}
The basic quantities that are normally taken into account for evaluating the detectability at VHE of 
a given $\gamma$-ray source with IACTs are the sensitivity\footnote{ 
The sensitivity $S$ of an IACT in a given energy interval $\Delta E$ is defined 
as the minimum flux of $\gamma$-ray events in $\Delta E$ (per unit time and area) 
that, in a given observation time, results in a statistically significant excess 
above the isotropic background of cosmic-ray initiated showers. When comparing 
different instruments, it is most often assumed that the source is point-like, 
and that its energy spectrum is a pure power-law of spectral index of $-2.6$ 
(which is the Crab Nebula index around $1$~TeV). A common sensitivity unit for 
different IACTs is the flux that will be measured with a significance ($\sigma$) greater
than $5$ in $50$~hours of observations (i.e. $S_{5\sigma,50\mbox{h}}$). The flux is 
typically expressed as a fraction of the Crab Nebula flux (Crab Units, CU).} 
of the instrument and the flux level of the source. However, these quantities are useful for 
detection considerations under the hypothesis of a steady $\gamma$-ray emission. In case of 
transient $\gamma$-ray events, whose flux is strongly time dependent, a different
approach is therefore needed. In this respect, a more useful quantity that can be considered is 
the significance of the observation ($\sigma$) as a function of time, 
provided an emission model for the transient source. In this way, in fact, it is possible to evaluate whether the 
typical detection condition $\sigma>5$ is achieved or not (in a certain energy interval and for different observational conditions).

The commonly used definition of $\sigma$ for IACT observations is given in Eq.~17 of~\cite{li83}:
\begin{equation}
\scriptsize{\sigma(N_{on},N_{off},\alpha) = 
\sqrt{2} \cdot \sqrt{N_{on} \ln \left[ \frac{(1+\alpha)N_{on}}{\alpha(N_{on}+N_{off})} \right] + N_{off} \ln \left[ \frac{(1+\alpha)N_{off}}{N_{on}+N_{off}} \right]}}~,
\label{eq1}
\end{equation}
where $N_{on}$ and $N_{off}$ are the number of events in the signal region
of the $ON$ and $OFF$\footnote{
In the IACT observations, the $OFF$ data set is needful to estimate the amount of irreducible 
background events $N_{bkg}$ in the $ON$ data set. The number of $\gamma$-ray excess events in
the $ON$ data set is given by $N_{\gamma} = N_{on} - \alpha N_{off} = N_{on} - N_{bkg}$.}
data sets, and $\alpha$ is the $ON$--$OFF$ normalization factor expressed (for real observations)
as the ratio between the effective time of the $ON$ and $OFF$ data sets, which implies that the expected amount 
of irreducible background in the $ON$ data set is $N_{bkg}=\alpha N_{off}$. 

Since $N_{on}$ and $N_{off}$ refer to a given energy interval $\Delta E$
and are functions of time, the significance is energy and time dependent. In addition, in case 
of transient event observations, the starting time of observation T$_{\mbox{s}}=$T$_{0}+\Delta$T 
(where T$_{0}$ is the time of the transient event burst) must be taken into account to define 
the initial time at which the source emission must be considered. 

In order to evaluate how the significance of a given short time-scale transient event observation 
evolves with time, in a given energy interval $\Delta E$\footnote{
In this work we consider the energy bins $\Delta E^{i}$ defined in~\cite{al12,be13},
i.e. 5 logarithmic energy bins per decade in the $10$\,GeV~--~$100$\,TeV band.
Hereafter, generic energy intervals are defined as 
$\Delta E \equiv \Delta E^{j,k}=\sum_{i=j}^{k}\Delta E^{i}$ (with $1<j<20$, $1<k<20$, $j \leq k$).
}, 
and for a given starting time of observation 
T$_{\mbox{s}}$, the following quantities must be taken into account:
\begin{itemize}
\item The number of $\gamma$-ray excess events from the transient source as a function of time,
in the energy bin $\Delta E^{i}$. This quantity can be calculated as
\begin{equation}
N_{\gamma}^{[\Delta E^{i},\mbox{T}_{\mbox{s}}]}(\widetilde{t})=
A_{eff}^{\Delta E^{i}} \times \int_{\mbox{T}_{\mbox{s}}}^{\widetilde{t}} \int_{\Delta E^{i}} \frac{\mbox{d}\Phi}{\mbox{d}E}(E,t)\mbox{d}E\mbox{d}t~,
\label{eq2}
\end{equation}
where $A_{eff}^{\Delta E^{i}}$\footnote{
The $\gamma$-ray effective collection area values $A_{eff}^{\Delta E^{i}}$ are
obtained from Monte Carlo (MC) simulations and depend on the simulated energy spectrum.
The values reported in Tab.~\ref{tab1} for MAGIC and CTA are calculated assuming 
a spectral power-law slope close to $-2.6$. 
In this work we use those values, although the transient event spectrum 
can have, locally, slopes much different from $-2.6$ (see e.g. Fig.~\ref{fig1}). However, 
we estimated (with MAGIC MC simulations) that for tentative spectral 
slopes down to $-5$ the obtained effective collection area values 
are within $15\%$ of the values given in Tab.~\ref{tab1}, in the considered energy bins. 
In order to cope with this issue, and to take into 
account the typical assumed systematic error of the effective collection area (around $30\%$), 
we consider a systematic error of $50\%$ on the effective collection area values
reported in Tab.~\ref{tab1}, and propagate that uncertainty in the calculation of the number 
of $\gamma$-ray excess events provided by Eq.~\ref{eq2}.}
is the (average) effective collection area
of the instrument in the $i$-th energy bin, and d$\Phi$/d$E$ is the differential energy spectrum model of the given transient event emission 
as a function of energy and time. The effect of the $\gamma$-ray attenuation by pair 
production with EBL photons~\cite{do11} must be taken into account in the spectrum model.
\item The number of background events as a function of time, in the energy bin $\Delta E^{i}$, given by
\begin{equation}
N_{bkg}^{[\Delta E^{i},\mbox{T}_{\mbox{s}}]}(\widetilde{t})=\frac{\mbox{d}N_{bkg}^{\Delta E{i}}}{\mbox{d}t} \cdot (\widetilde{t}-\mbox{T}_{\mbox{s}})~,
\label{eq3}
\end{equation}
where d$N_{bkg}^{\Delta E{i}}/$d$t$ is the background rate of the instrument
in the $i$-th energy bin. In the present work, this quantity is assumed to be independent of time and 
of the telescope azimuthal pointing. 
\end{itemize}
The significance of a given transient event observation as a function of time, in a given 
energy interval $\Delta E$, and for a given starting time of observation T$_{\mbox{s}}$, 
is thus given by
\begin{equation}
\scriptsize{\sigma^{[\Delta E,\mbox{T}_{\mbox{s}}]}(\widetilde{t})=
\sigma(\sum_{i=j}^{k}[N_{\gamma}^{[\Delta E^{i},\mbox{T}_{\mbox{s}}]}(\widetilde{t})+N_{bkg}^{[\Delta E^{i},\mbox{T}_{\mbox{s}}]}(\widetilde{t})],\alpha^{-1} \sum_{i=j}^{k}N_{bkg}^{[\Delta E^{i},\mbox{T}_{\mbox{s}}]}(\widetilde{t}),\alpha)}~,
\label{eq4}
\end{equation}
where $N_{\gamma}^{[\Delta E^{i},\mbox{T}_{\mbox{s}}]}$ and $N_{bkg}^{[\Delta E^{i},\mbox{T}_{\mbox{s}}]}$ 
are defined in Eq.~\ref{eq2} and Eq.~\ref{eq3}, respectively, and $\alpha$ is equal to $1$ 
for MAGIC~\cite{al12} and $0.2$ for CTA~\cite{be13}.

In Tab.~\ref{tab1}, the main quantities needed for the calculation of the significance
as a function of time (provided an emission model for the transient source) for 
MAGIC~\cite{al12} and CTA (candidate array I)~\cite{be13}, in $5$ logarithmic energy bins between 
$10^{1.6}$\,GeV and $10^{2.6}$\,GeV\footnote{
We restrict our attention to these energy bins because transient events are
foreseen to have VHE emission that rapidly vanish at energies of a few hundreds of GeV,
due the $\gamma$-ray EBL absorption.},
are shown. All quantities refer to point-like source observations.
For completeness, the differential sensitivities $S_{5\sigma,50\mbox{h}}$ of the MAGIC telescopes and 
CTA (candidate array I) are also reported. 
\begin{table}[t]
\centering
\begin{tabular}{c c c c c}
\hline
\small $E_{min}$ & \small $E_{max}$ & \small $bkg$-rate   & \small A$_{eff}$ & \small $S_{5\sigma,50\mbox{h}}$\\ 
\small [GeV]     & \small [GeV]     & \small [min$^{-1}$] & \small [m$^{2}$] & \small [\%CU]\\
\hline
\multicolumn{5}{c}{\small{MAGIC}}\\
\hline
\small 39.8      & \small 63.1      & \small 1.61      & \small 673       & \small 59.01~(39.99)\\
\small 63.1      & \small 100       & \small 3.01      & \small 5914      & \small 16.58~(10.52)\\
\small 100       & \small 158.5     & \small 2.04      & \small 24334     & \small 6.93~(3.65)\\
\small 158.5     & \small 251.2     & \small 0.61      & \small 31903     & \small 6.48~(2.70)\\ 
\small 251.5     & \small 398.1     & \small 0.13      & \small 33302     & \small 7.39~(2.20)\\
\hline
\multicolumn{5}{c}{\small{CTA (candidate array I)}}\\
\hline
\small 39.8      & \small 63.1      & \small 2.45      & \small 7719      & \small 3.11\\
\small 63.1      & \small 100       & \small 0.90      & \small 15233     & \small 1.70\\
\small 100       & \small 158.5     & \small 0.68      & \small 40451     & \small 1.12\\
\small 158.5     & \small 251.2     & \small 0.04      & \small 32501     & \small 0.69\\
\small 251.5     & \small 398.1     & \small 0.02      & \small 58559     & \small 0.56\\
\hline
\end{tabular}
\caption{
Main quantities needed for the detection prospect method for the MAGIC 
telescopes~\cite{al12} and CTA (candidate array I)~\cite{be13}.
The values are given for $5$ logarithmic energy bins between $10^{1.6}$\,GeV and $10^{2.6}$\,GeV. 
The significance $\sigma$ used for the computation of the sensitivity $S_{5\sigma,50\mbox{h}}$
is defined in Eq.~$17$ of~\cite{li83} (see Eq.~\ref{eq1}), with $\alpha$ equal to $1$ for MAGIC and $0.2$
for CTA. The MAGIC differential sensitivities reported in parenthesis are
obtained with significance defined as $\sigma=N_{\gamma}/\sqrt{N_{bkg}}$.
}
\label{tab1}
\end{table}

\section{Test case: GRB~090102}
\label{sec4}
\begin{figure}[t]
\centering
\includegraphics[width=0.4\textwidth]{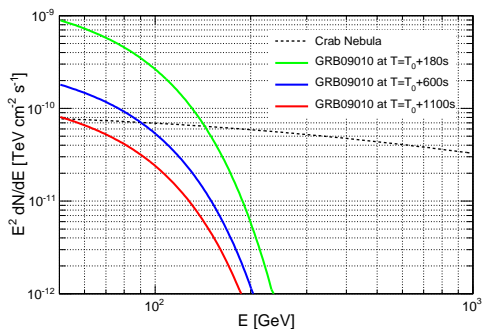}
\caption
{
The modeled SED emission of GRB~090102 from SSC mechanism at different times: $\mbox{T}_{0}+180$\,s (green line), 
$\mbox{T}_{0}+600$\,s (blue line), and $\mbox{T}_{0}+1100$\,s (red line). For comparison,
the Crab Nebula SED, as measured by MAGIC~\cite{al12}, is also drawn (dashed black line).
}
\label{fig1}
\end{figure}

As a test case for the detection prospect method presented in Sec.~\ref{sec3}, we consider the
GRB~090102 event. This GRB was detected and located by the \emph{Swift} satellite on January $2^{{\rm nd}}$, 
2009, at 02:55:45 UT~\cite{ma09}. The MAGIC-I telescope observed GRB~090102 after $\sim 1100$\,s from the 
event burst, deriving flux upper limits above $\sim 50$\,GeV~\cite{al13a}.
The prompt light curve was structured in four partially overlapping 
peaks for a total T$_{90}$ of $27.0 \pm 2.0$\,s. The moderate measured redshift of $z=1.547$ 
implies an isotropic energy value of $E_{iso} = 5.75 \times 10^{53}$\,erg. According to the relativistic
blast-wave model~\cite{zh01}, we use the Synchrotron Self-Compton (SSC) mechanism to derive the expected 
VHE emission during the afterglow in the IACT energy range.
The Spectral Energy Distribution (SED) of the event can be expressed as
\begin{equation}
E^2\frac{\mbox{d}\Phi}{\mbox{d}E}(E,t,z) = \phi_{0} \left(\frac{E}{\mathrm{1~TeV}}\right)^{\frac{2-p}{2}} \left(\frac{t}{\mathrm{1~s}}\right)^{\frac{10-9p}{8}} \mbox{e}^{-\tau(E,z)}~,
\label{eq6}
\end{equation}
where $\phi_{0}=0.78\times10^{-6}~\mathrm{TeV~cm^{-2}~s^{-1}}$ is the normalization constant at $1$\,TeV
and 1~s, $p=2.29$ is the index of the electrons power-law distribution~\cite{ge10}, and $\mbox{e}^{-\tau(E,z)}$ 
is the EBL absorption factor evaluated for $z=1.547$ using the model by~\cite{do11}. In Fig.~\ref{fig1} the modeled
SED of GRB~090102, at three different times after the event burst, is shown.

\begin{figure}[t]
\centering
\includegraphics[width=0.38\textwidth]{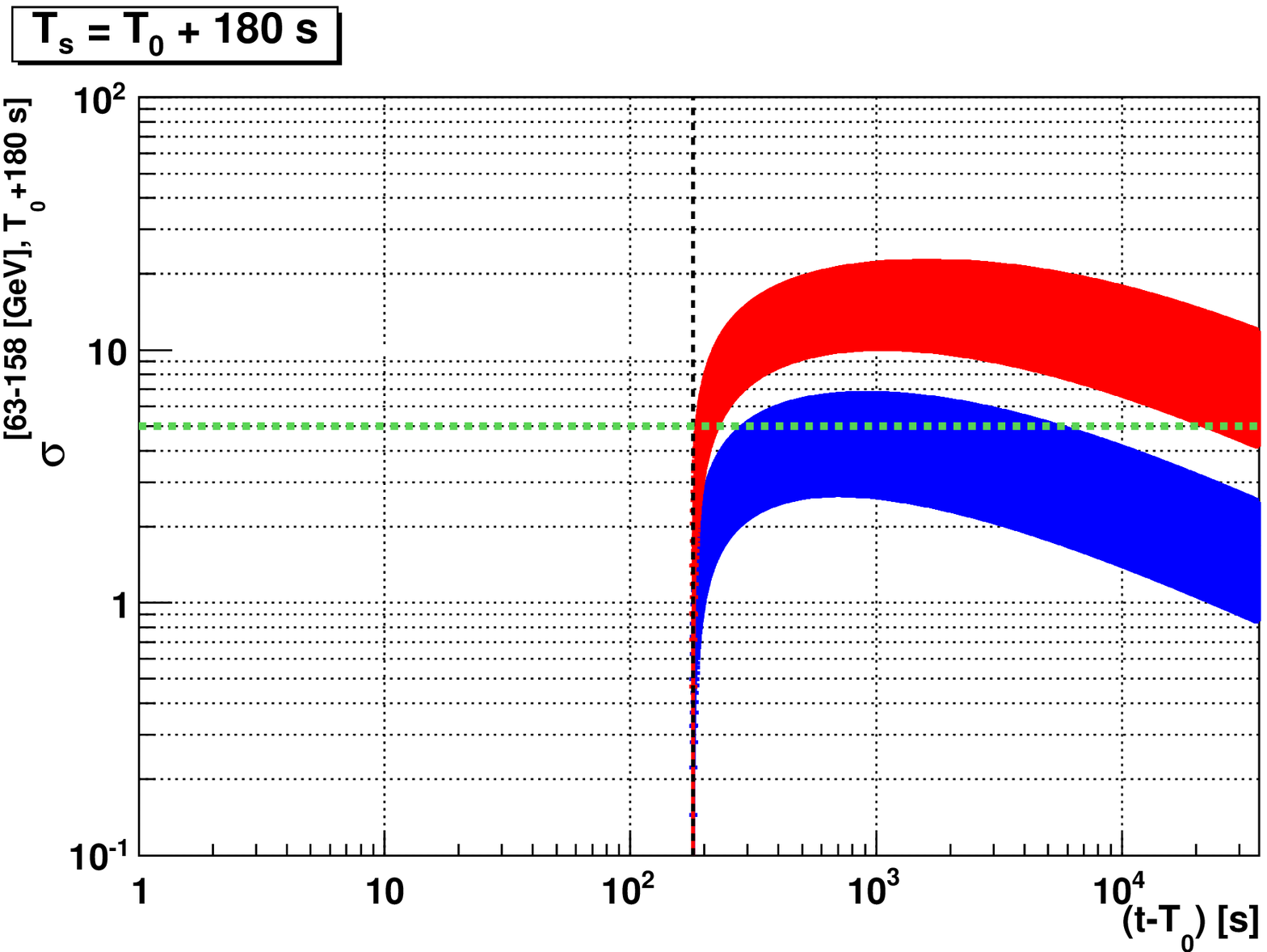}
\includegraphics[width=0.38\textwidth]{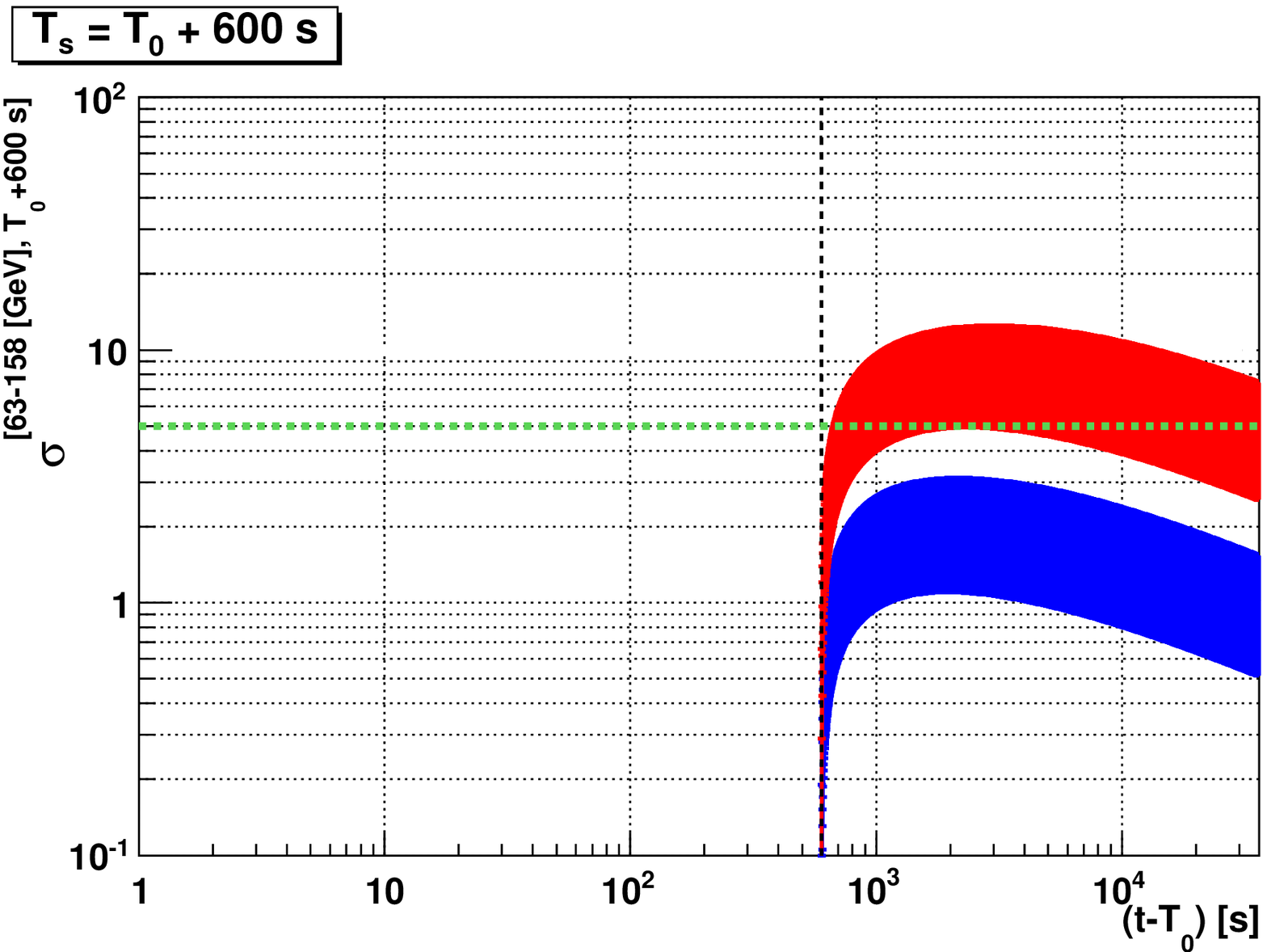}
\includegraphics[width=0.38\textwidth]{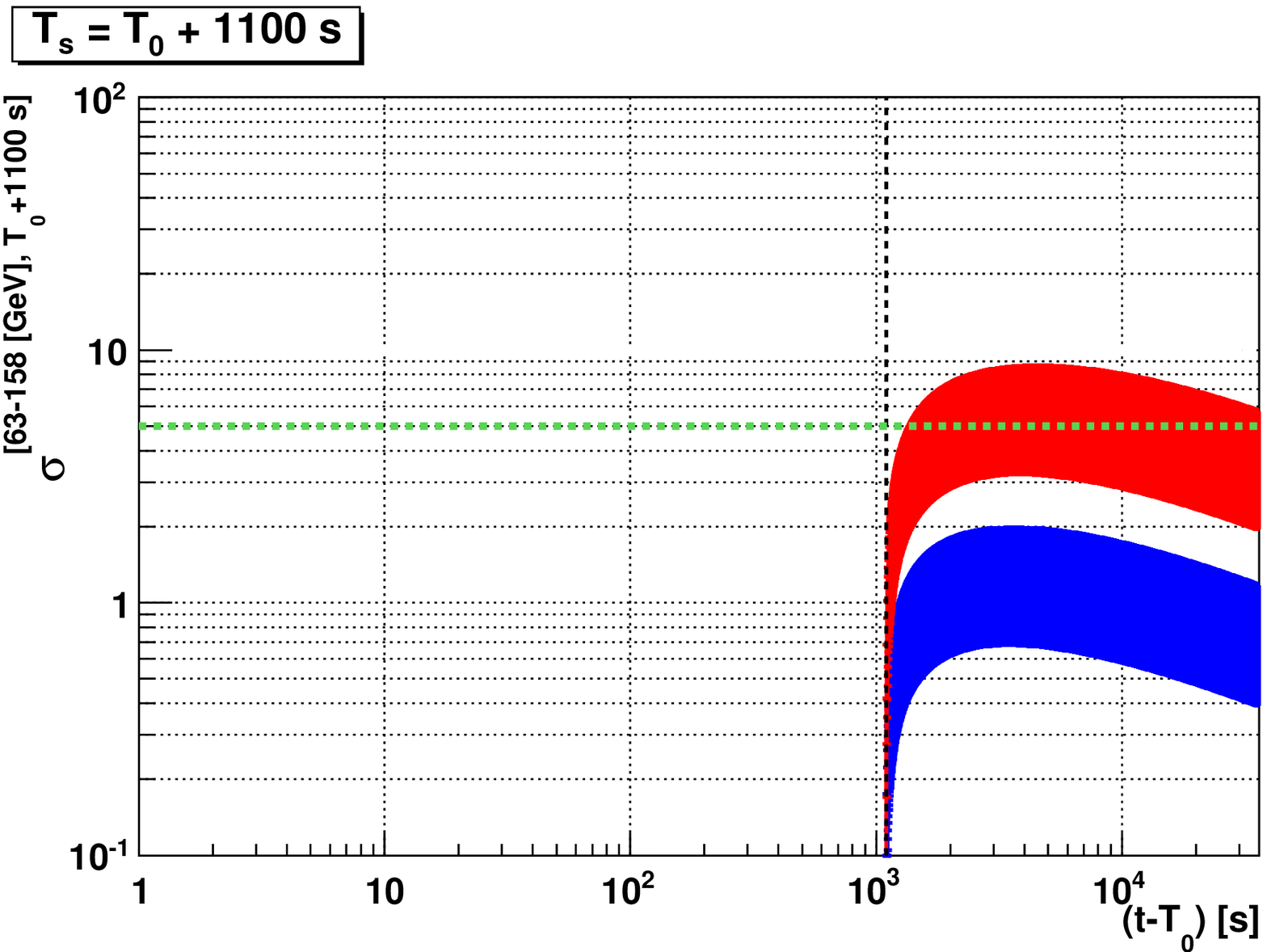}
\caption{
MAGIC (blue area) and CTA (candidate array I, red area) significance of GRB~090102 observation,
in the energy interval $63.5<E[\mbox{GeV}]<158.1$, for three different starting times of 
observation: T$_{\mbox{s}}=$T$_{0}+180$\,s (upper plot), T$_{\mbox{s}}=$T$_{0}+600$\,s
(middle plot) and T$_{\mbox{s}}=$T$_{0}+1100$\,s (lower plot). The green dashed horizontal
line represents the detection threshold $\sigma=5$. 
A systematic error of $50\%$ on the effective collection area values is taken into account
in the significance calculations.
}
\label{fig2}
\end{figure}

Using Eq.~\ref{eq1}, Eq.~\ref{eq4}, the quantities reported in Tab.~\ref{tab1}, and
the SED emission model defined in Eq.\ref{eq6}, we can estimate how the significance of the 
GRB~090102 observation would be with the MAGIC stereoscopic system and CTA as a function of time, 
for different observational conditions. 
In Fig.~\ref{fig2}, we present the achieved results in the energy interval $63.1<E~[\mbox{GeV}]<158.5$
(where, from our estimates, the GRB~090102 detection prospects turn out to be the most favourable) 
and for three different starting times of observation: T$_{\mbox{s}}=$T$_{0}+180,600,1100$\,s. 
A systematic error of $50\%$ on the effective collection area values is taken into account
in the significance calculations.

As expected, the starting time of observation T$_{\mbox{s}}$ (in case of SSC emission model) is a
crucial parameter: the earlier the IACT observation starts after the transient event burst, the higher is
the possibility to detect a $\gamma$-ray signal from the source. Furthermore, it is interesting to 
point out how the CTA performance would allow a significant detection of the event even up to T$_{\mbox{s}} \simeq T_{0}+1$\,ks, 
while, in case of MAGIC observation, the source would be detectable only for starting times of observation 
T$_{\mbox{s}}<$~T$_{0}+180$\,s.

\section{Conclusions}
\label{sec5}
One of the primary goals for current IACTs, like the MAGIC telescopes, and for future 
Cherenkov Telescope Array is to catch VHE signal from GRBs. In this contribution, we 
presented a method aimed at providing detection prospects for short time-scale transient 
events at VHE (provided their emission model), and considered the particular event GRB~090102 
as a test case.

Our estimates show that, for this particular event, MAGIC follow-up observations made within 
a couple of minutes from the event onset would have the potential to detect the VHE component 
or at least to derive constraining upper limits. In fact, the steep time decay of the source 
(as $t^{-1.1 \div 1.2}$) makes a MAGIC detection at later starting times (T$_{\mbox{s}}>$T$_{0}+200$\,s)
unlikely, while 
interesting prospects for an afterglow significant detection in the VHE domain at such later 
times are possible within the CTA context.

The possibility to extend our method to other classes of variable and transient sources is going 
to be investigated producing reliable detection prospects at VHE for the coming age of time domain 
astrophysics.


\end{document}